\documentclass{nature}

\usepackage{amsmath}    
\usepackage{graphicx}   
\usepackage{color}
\usepackage[10pt]{moresize}
\usepackage{graphics,graphicx}
\usepackage{amsfonts}
\usepackage{amssymb}
\usepackage{amscd}
\usepackage{amsmath}
\usepackage{enumerate}
\usepackage{epsfig}
\usepackage{subfigure}

\let\oldcaption\caption
\renewcommand{\caption}{\sffamily \oldcaption}
\newcommand{\bra}[1]{\mbox{$\left\langle #1 \right|$}}
\newcommand{\ket}[1]{\mbox{$\left| #1 \right\rangle$}}

\begin{document}

\title{Security of quantum key distribution with multiphoton components}

\author{Hua-Lei Yin$^{1,2,\ast}$, Yao Fu$^{1,2,\ast}$, Yingqiu Mao$^{1,2}$ \& Zeng-Bing Chen$^{1,2}$ }

\maketitle

\begin{affiliations}
\item
Hefei National Laboratory for Physical Sciences at Microscale and Department of Modern Physics, University of Science and Technology of China, Hefei, Anhui 230026, China
\item
The CAS Center for Excellence in QIQP and the Synergetic Innovation Center for QIQP, University of Science and Technology of China, Hefei, Anhui 230026, China \\
$^\ast$ These authors contributed equally to this work.\\
Correspondence and requests for materials should be addressed to H.-L.Y. (email: hlyin@mail.ustc.edu.cn) or Z.-B.C. (email: zbchen@ustc.edu.cn)
\end{affiliations}

\baselineskip24pt

\maketitle

\begin{abstract}
Most qubit-based quantum key distribution (QKD) protocols extract the secure key merely from single-photon component of the attenuated lasers.
However, with the Scarani-Acin-Ribordy-Gisin 2004 (SARG04) QKD protocol, the unconditionally secure key can be extracted from the two-photon component by modifying the classical post-processing procedure in the BB84 protocol.
Employing the merits of SARG04 QKD protocol and six-state preparation, one can extract secure key from the components of single photon up to four photons. In this paper, we provide the exact relations between the secure key rate and the bit error rate in a six-state SARG04 protocol with single-photon, two-photon, three-photon, and four-photon sources. By restricting the mutual information between the phase error and bit error, we obtain a higher secure bit error rate threshold of the multiphoton components than previous works. Besides, we compare the performances of the six-state SARG04 with other prepare-and-measure QKD protocols using decoy states.
\end{abstract}

Quantum key distribution (QKD)\cite{BB_84,e91quantum} offers information-theoretic security for two authorized users, Alice and Bob, when communicating secret information along an insecure quantum channel, while the laws of quantum mechanics bound the behavior of an eavesdropper\cite{lo1999unconditional,shor2000simple,Kraus:2005:Lower,Tomamichel:2011}
. Since its introduction in 1984 by Bennett and Brassard\cite{BB_84}, QKD has experienced great advances both theoretically \cite{RevModPhys:02:Gisin,wang2007quantum,RevModPhys:09:Scarani,Weedbrook:2012:CV,lo2014secureQKD,ekert:2014:ultimate} and experimentally \cite{takesue2007quantum,wang:2012,Liu:2013:Experimental,Tang2014MDI200km,korzh2015NatPhot307km,Tang2016MDInet}, and has become the most mature quantum information technology for commercial use\cite{frohlich2013QNetwork}. The study of QKD today is driven by the necessity to close the gap between its theory and practice, as experimental systems tend to differ remarkably from their simplified mathematical models, and any of these deviations may open doors to new attacks from Eve to compromise security. Some of Eve's eavesdropping techniques include simple individual attacks and Trojan-horse attacks, which one can overcome by investigating the bounds of information leakage in different scenarios and apply the suitable amount of privacy amplification to obtain the final secure key\cite{lucamarini2015PRXTHA}. Other side-channel attacks, such as detector blinding attacks\cite{Lydersen:BrightAttack:2010} and time-shift attack\cite{Zhao:2007:Quantum} that base on specific device imperfections, require more complicated QKD settings than the original BB84 to retrieve security again. Hence the measurement-device-independent (MDI) QKD\cite{Lo:2012:MDI,Braunstein:2012:Side,Ma:Statistical:2012,yin2014long,curty:2014:finite,wang2014simulating,yin2014measurement,zhou2015making,Xu:2015:Measurement,Fu:2016:Long} and device-independent (DI) QKD\cite{Acin:2007:DIQKD,Gisin2010DIQKD,Vazirani2014FullDIQKD} were developed to combat these experimental flaws.

Compared with the entanglement-based QKD protocols, prepare-and-measure QKD protocols are widely studied.
The photon-number distribution of weak coherent states is Poisson distribution, which contains a fraction of multiphoton components. However, exploiting photon-added coherent states\cite{Agarwal:1991:Agarwal}, one can acquire large probabilities of single-photon, two-photon, three-photon or four-photon component.
For the BB84 protocol, the single-photon source is usually replaced by weak coherent states, which suffer from the photon number splitting (PNS) attack\cite{Brassard2000PNS}.
The PNS attack, in which Eve blocks all single photon pulses and splits multiphoton pulses, results from the experimental variation of replacing the single photon sources from the original BB84 protocol with practical attenuated lasers.
In this situation, Eve would forward some portion of multiphoton pulses to Bob through a lossless channel while keeping the rest to herself in the quantum memory\cite{azuma2015all,abruzzo2014measurement,mirza2013single}, and measure her photons after receiving the basis reconciliation information obtained via Alice and Bob's public communication. The security basis of QKD provided by single photon pulses was guaranteed by the no-cloning theorem\cite{wootters1982NoCloning}, and thus this attack was regarded as a major threat to QKD and has been extensively studied\cite{Brassard2000PNS}.
Two major counter methods have been proposed.
One is the decoy state method\cite{Hwang:2003:Quantum,Lo:2005:Decoy,Wang:2005:Beating}, which is a powerful method devised to analyze rigorously the extractible secret key rate from the single-photon component of signal states, though its implementations would differ slightly from the prepare-and-measure setup\cite{peng2007ExpDecoyQKD,Rosenberg:2007:Long,Schmitt:2007:Exp}.
To overcome this attack at a protocol level, the SARG04 QKD protocol\cite{SARG04:2004:Quantum}, which differs from the BB84 only in the classical post-processing part\cite{acin2004SARG04PRA,Branciard2005SARG04}, was proposed. In the SARG04 protocol, the reconciliation information is a pair of nonorthogonal states, which cannot be perfectly distinguished and can be able to address the PNS attack\cite{SARG04:2004:Quantum}.
Subsequently, this prepare-and-measure SARG04 protocol was further investigated and ingeniously converted to an unconditionally secure entanglement distillation protocol (EDP) by Tamaki and Lo\cite{Tamaki:2006:SARGtwophoton}, who showed that by exploiting the same arguments of Shor and Preskill\cite{shor2000simple}, SARG04 protocol possesses the unique ability to extract the secure key from not only the single-photon component but also two-photon component\cite{Tamaki:2006:SARGtwophoton,Fung2006SARG04}.
This opens the interesting question that under certain modifications of the original BB84 protocol, how the secret key can be extracted from multiphoton states. The SARG04 protocol has been widely investigated in theories\cite{acin2004SARG04PRA,Branciard2005SARG04,koashi2005security ,Tamaki:2006:SARGtwophoton,Fung2006SARG04,fang2010passive,ali2010fiber,ali2012practical} and experiments\cite{stucki2011long,jeong2011effects}.
Similarly to the MDIQKD protocol, which was proposed to make BB84 protocol naturally immune to all side-channel attacks on detectors, the SARG04 protocol in MDI setting has been considered likewise\cite{mizutani2014measurement}.
Also, with the advantage of secure key generation from two-photon component, recently a nonorthogonal state encoding method of SARG04 has been successfully applied to circumvent the forging attack of quantum digital signature with insecure quantum channel\cite{Yin:2016:QDS}.
A generalized SARG04 with six states (six-state SARG04) protocol has been analyzed in ref. 51, which showed that one could extract the secure key from the single-photon component to up to four-photon component. However, without the explicit relations between the phase error rate and bit error rate of the six-state SARG04 protocol\cite{Tamaki:2006:SARGtwophoton}, one cannot acquire the exact secure key rate.

Here, we generalize the EDP of ref. 51 to acquire the exact relationships between the phase error rate and bit error rate of single-photon, two-photon, three-photon, and four-photon components in the six-state SARG04 protocol. Furthermore, we carefully analyze the mutual information between phase error and bit error, and discover that the mutual information of two-photon component in a four-state SARG04 protocol and that of three-photon and four-photon components in a six-state SARG04 protocol are not zero, so the secure bit error rate threshold of those cases are higher than the results of previous works\cite{Tamaki:2006:SARGtwophoton}. Finally, we perform a numerical simulation to study the performance of six-state SARG04 with weak coherent states in an infinite decoy states setting.
Also, we compare the performance of six-state SARG04 and other prepare-and-measure QKD protocols, i.e., BB84\cite{BB_84,Lo:2005:Decoy}, four-state SARG04\cite{SARG04:2004:Quantum,Fung2006SARG04}, and round-robin differential phase-shift (RRDPS) QKD protocols\cite{sasaki2014practical,zhang2015round,Yin:2016:Detector} in the same situation.

\section*{Results}
\textbf{Six-state SARG04 QKD protocol.} In this section, we introduce the six-state SARG04 QKD protocol with $\nu$-photon ($\nu\in\{1,2,3,4\}$) source.
In this protocol, there are six polarization encoding quantum states, $\ket{H}$, $\ket{V}$, $\ket{\pm}=(\ket{H}\pm\ket{V})/\sqrt{2}$, $\ket{R}=(\ket{H}+i\ket{V})/\sqrt{2}$, and $\ket{L}=(\ket{H}-i\ket{V})/\sqrt{2}$. The six states are then arranged into twelve sets $\{\ket{H},\ket{-}\}$, $\{\ket{-},\ket{V}\}$, $\{\ket{V},\ket{+}\}$, $\{\ket{+},\ket{H}\}$, $\{\ket{H},\ket{R}\}$, $\{\ket{R},\ket{V}\}$, $\{\ket{V},\ket{L}\}$, $\{\ket{L},\ket{H}\}$, $\{\ket{R},\ket{-}\}$, $\{\ket{-},\ket{L}\}$, $\{\ket{L},\ket{+}\}$, $\{\ket{+},\ket{R}\}$, where the first and second states of each set correspond to logic 0 and 1, respectively. The steps of the six-state SARG04 QKD protocol with a $\nu$-photon source are outlined as follows.
Alice sends a series of signals to Bob. Each pulse is chosen randomly from the twelve sets listed above, and Alice randomly sends one state from each set to Bob through the insecure quantum channel.
Bob randomly measures the incoming bit strings with one of the three bases, $Z$, $X$, and $Y$. Afterwards, he exploits
an authenticated classical channel to announce to Alice the situations where he did not register any click at his detection unit, and both of them discard these signals.
Alice reveals to Bob the sets on which she encodes her information.	
Bob then compares his measurement results with Alice's set information. If Bob's measurement result is orthogonal to one of the states in the set, he concludes that
the other state has been sent, which represents a conclusive result; he concludes an inconclusive result otherwise. He discards all the inconclusive results and broadcasts to Alice which of his results are conclusive.
Alice selects randomly a portion of her remaining signals and announces them to Bob, and Bob calculates the bit error rate  to test for eavesdroppers. If the bit error rate is much higher than the upper bound, they abort the protocol.
They perform error correction and privacy amplification on the remaining bit string to obtain the shared secret key.

\textbf{A virtual EDP-based six-state SARG04 protocol.}
To estimate phase error for privacy amplification, one can construct an equivalent EDP version of the six-state SARG04 protocol. First, we introduce some notations. $\{\ket{0_x},\ket{1_x}\}$ and $\{\ket{0_z},\ket{1_z}\}$ are the eigenstates for $X$ and $Z$ basis, respectively. $R=\cos(\pi/4)\textrm{I}+\sin(\pi/4)(\ket{1_x}\bra{0_x}-\ket{0_x}\bra{1_x})$; $T_0=\textrm{I}$ is an identity operation; $T_1=\cos(\pi/4)\textrm{I}-i\sin(\pi/4)\frac{Z+X}{\sqrt{2}}$ is a $\pi/2$ rotation around the $\frac{Z+X}{\sqrt{2}}$ axis; $T_2=\cos(\pi/4)\textrm{I}-i\sin(\pi/4)\frac{Z-X}{\sqrt{2}}$ is a $\pi/2$ rotation around the $\frac{Z-X}{\sqrt{2}}$ axis. In the EDP-$\nu$ protocol, Alice prepares many pairs of qubits in the state $\ket{\Psi^{(\nu)}}_{AB}=(\ket{0_z}_{A}\ket{\varphi_0}_{B}^{\otimes\nu}+\ket{1_z}_{A}\ket{\varphi_1}_{B}^{\otimes\nu})/\sqrt{2}$,
where $\ket{\varphi_j}=\cos(\pi/8)\ket{0_x}+(-1)^j\sin(\pi/8)\ket{1_x}$ ($j\in\{0,1\}$). She then randomly rotates $T_{l}R^k$ and sends system $\textrm{B}$ to Bob, where $l\in\{0,1,2\}$ and $k\in\{0,1,2,3\}$.
Upon receiving the qubits, Bob first applies a random reverse rotation $ R^{-k'}T_{l'}^{-1}$, before performing a filtering operation defined by a Klaus operator $F=\sin(\pi/8)\ket{0_x}\bra{0_x}+\cos(\pi/8)\ket{1_x}\bra{1_x}$.
Then Alice and Bob would compare their indices ${k,l}$ and ${k',l' }$ via public communication, and keep the qubit pairs with $k=k'$ and $l=l'$ when Bob's filtering operation is successful.
They then choose some states randomly as test bits and measure them in the $Z$ basis, and compare their results publicly to estimate the bit error rate and the information acquired by Eve.
Finally, they utilize the corresponding Calderbank-Shor-Steane (CSS) code to correct the bit and phase errors\cite{shor2000simple}, and perform a final $Z$ measurement on their qubits to obtain the secure key.

The six-state SARG04 QKD protocol is equivalent to the EDP-based six-state SARG04 QKD protocol, except for the only difference, a $\pi/4$ total rotation around $Y$ basis.
By analyzing the virtual EDP-based six-state SARG04 QKD protocol, we give the exact phase error rate formula, whose detailed analysis is provided in the Methods section.
For the case with a single-photon source, we have
\begin{equation} \label{qubit}
\begin{aligned}
e_{p}&=\frac{3}{2}e_{b},~~ a=\frac{3}{4}e_{b},
\end{aligned}
\end{equation}
where $e_{p}$ and $e_{b}$ are the phase error rate and bit error rate, respectively. $a$ is the probability that both bit flip and phase shift occur, which restricts the mutual information between phase error and bit error. For the case of a two-photon source, the relationship can be given by
\begin{equation} \label{qubit}
\begin{aligned}
e_{p}&=\frac{2-\sqrt{2}}{4}+\frac{3}{2\sqrt{2}}e_{b},~~a=\frac{4+\sqrt{2}}{8}e_{b}.
\end{aligned}
\end{equation}
For the case with a three-photon source, the error rates can be written as
\begin{equation} \label{qubit}
\begin{aligned}
e_{p}&=\frac{1}{4}+\frac{3}{4}e_{b},~~ \frac{1}{2}e_{b}\leq a\leq\frac{3}{4}e_{b}.
\end{aligned}
\end{equation}
For the case with a four-photon source, the error rates are calculated by
\begin{equation} \label{qubit}
\begin{aligned}
e_{p}&=\min_{x}\left\{xe_{b}+f(x)\right\}, \forall x, ~~\frac{2-\sqrt{2}}{4}e_{b}\leq a\leq\frac{4+\sqrt{2}}{8}e_{b},
\end{aligned}
\end{equation}
where
\begin{equation} \label{qubit}
\begin{aligned}
f(x)=\frac{6-4x+\sqrt{6-12\sqrt{2}x+16x^2}}{12}.
\end{aligned}
\end{equation}

Now we reexamine the four-state SARG04 QKD protocol\cite{Tamaki:2006:SARGtwophoton,Fung2006SARG04}, and we find that the mutual information between bit error and phase error of a two-photon source is not zero.
The expression can be given by
\begin{equation} \label{qubit}
\begin{aligned}
e_{p}&=\min_{x}\left\{xe_{b}+f(x)\right\}, \forall x, ~~\frac{2-\sqrt{2}}{4}e_{b}\leq a\leq\frac{2+\sqrt{2}}{4}e_{b}.
\end{aligned}
\end{equation}
where
\begin{equation} \label{qubit}
\begin{aligned}
f(x)=\frac{3-2x+\sqrt{6-6\sqrt{2}x+4x^2}}{6},
\end{aligned}
\end{equation}
from which we can see that this phase error rate formula is the same as the result in ref. 51.

\begin{figure}
\centering
\resizebox{10cm}{!}{\includegraphics{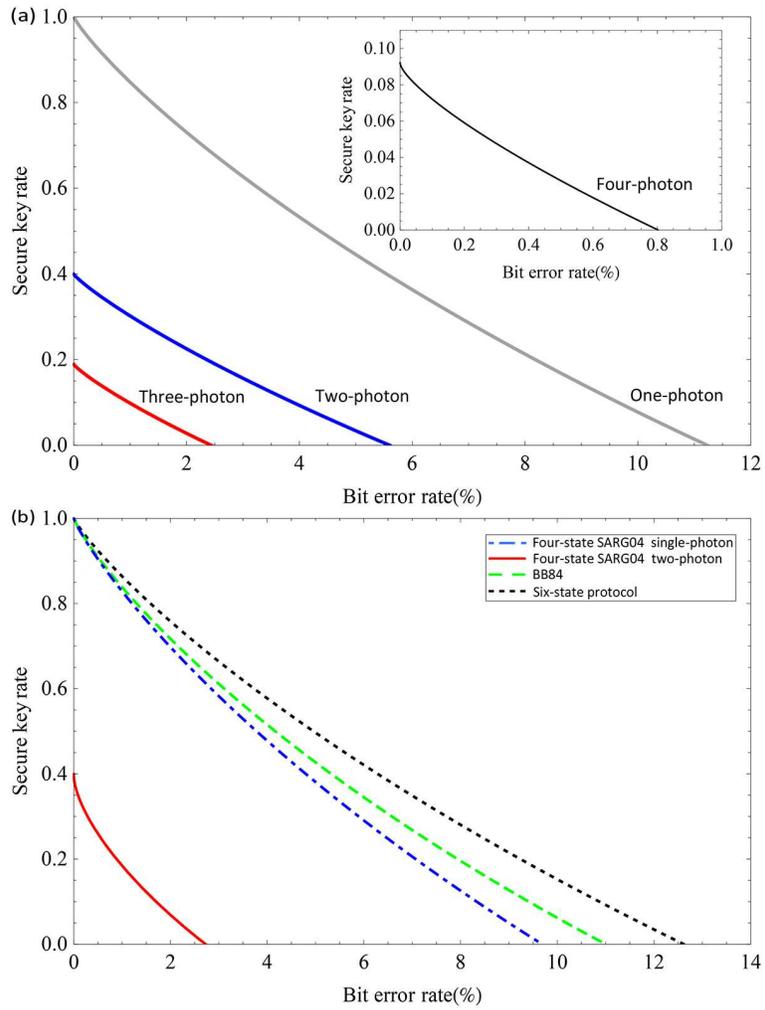}}
\caption{Plot of secure key rates versus bit error rates for (\textbf{a}) six-state SARG04 protocol with one-photon, two-photon, three-photon, and four-photon sources, and (\textbf{b}) BB84 protocol, six-state, and four-state SARG04 protocol for comparison.}
\label{f1}
\end{figure}

The secure key rate of the EDP-based QKD using one-way classical communication can be given by\cite{Fung2006SARG04}
\begin{equation} \label{qubit}
\begin{aligned}
r=1-H(e_{b})-H(e_{p}|e_{b}),
\end{aligned}
\end{equation}
where $H(x)=-x\log_{2}(x)-(1-x)\log_{2}(1-x)$ is the binary Shannon
entropy, $H(e_{p}|e_{b})$ is the conditional Shannon entropy function shown in the Methods section.
We calculate the secure key rates versus the bit error rates for the six-state SARG04 QKD protocol with single-photon, two-photon, three-photon and four-photon sources, as shown in Fig. \ref{f1}.
For comparison, we also calculate the secure key rate versus bit error rate for BB84 protocol\cite{shor2000simple}, six-state protocol\cite{lo2001proof}, and four-state SARG04 QKD protocol\cite{Tamaki:2006:SARGtwophoton}.
For the six-state SARG04 QKD protocol, one can extract the secure key rate from $\nu$-photon component when the bit error rate is no larger than $11.235\%$ (with $\nu=1$), $5.602\%$ (with $\nu=2$), $2.438\%$ (with $\nu=3$), and $0.802\%$ (with $\nu=4$).
For the four-state SARG04 QKD protocol with our calculation, one can extract the secure key rate from two-photon component when the bit error rate is lower than $2.726\%$. We can see that the bit error rate thresholds of single-photon and two-photon in the six-state SARG04 QKD protocol are the same with the results in ref. 51, while the bit error rate thresholds of three-photon and four-photon in the six-state SARG04 QKD protocol and two-photon in the four-state SARG04 QKD protocol are larger than the results in ref. 51.
If we neglect the mutual information between phase error and bit error, the bit error rate thresholds of three-photon (2.370\%) and four-photon (0.788\%) in the six-state SARG04 QKD protocol and two-photon (2.710\%) in the four-state SARG04 QKD protocol are the same with the results in ref. 51.

\begin{figure}
\centering
\resizebox{10cm}{!}{\includegraphics{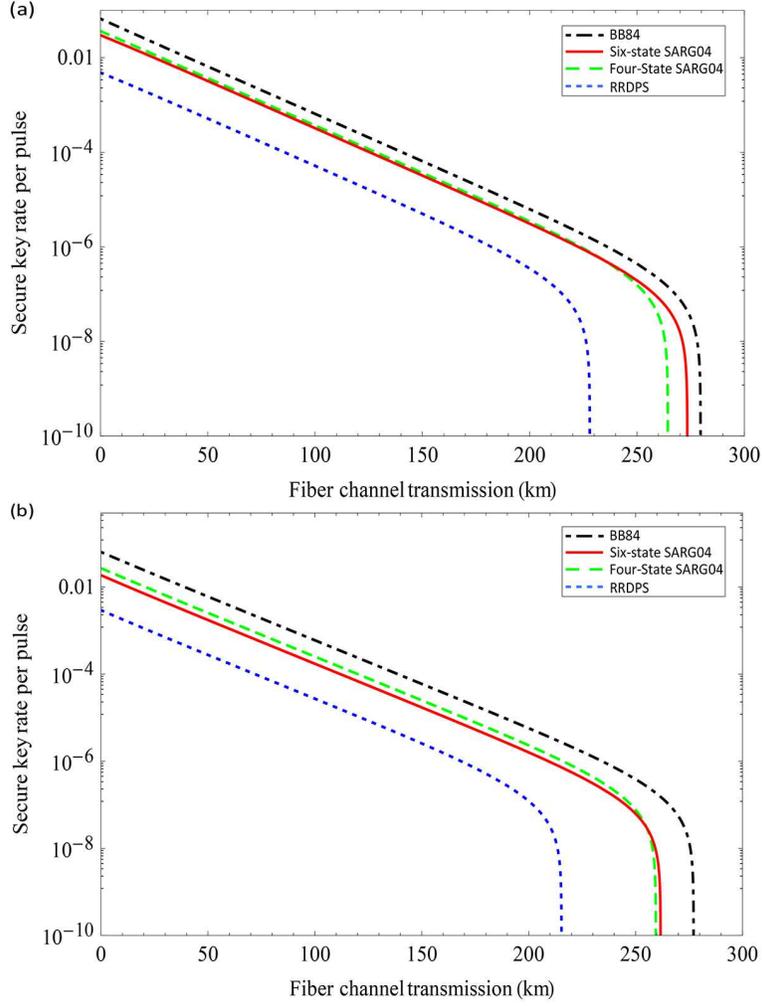}}
\caption{(\textbf{a}) Plot of secure key rate versus fibre channel transmission for various QKD protocols with infinite decoy states. For each transmission loss, we
optimize the intensity of signal states. For comparison, we set the number of pulses of each block $L=10$, since the phase error rate is $n/(L-1)$ for $n$-photon in the RRDPS QKD protocol\cite{sasaki2014practical}. It means that we can extract the secure key from single-photon, two-photon, three-photon and four-photon components for RRDPS QKD protocol. (\textbf{b}) Plot of secure key rate versus fibre channel transmission for various QKD protocols with one weak decoy state and vacuum state. }
\label{f2}
\end{figure}

For the phase randomized weak coherent state sources\cite{Tang:2013:source}, we study the secure key rate with infinite decoy states\cite{Hwang:2003:Quantum,Lo:2005:Decoy,Wang:2005:Beating}, which can be given by
\begin{equation} \label{qubit}
\begin{aligned}
R=&Q_{0}+Q_{1}[1-H(e_{p_{1}}|e_{b_{1}})]+Q_{2}[1-H(e_{p_{2}}|e_{b_{2}})]+Q_{3}[1-H(e_{p_{3}}|e_{b_{3}})]\\
&+Q_{4}[1-H(e_{p_{4}}|e_{b_{4}})]-Q_{\mu}fH(E_{\mu}),
\end{aligned}
\end{equation}
where $Q_{n}$ is the gain of the $n$-photon signal states which can be estimated by the decoy-state method; $e_{p_{n}}$ ($e_{b_{n}}$) is the phase (bit) error for the $n$-photon state; $Q_{\mu}$ and $E_{\mu}$ are, respectively, the total gain and quantum bit error rate under signal states with $\mu$ intensity, and can both be acquired directly through the experiment.
We execute a numerical simulation to study its performance, as shown in Fig. \ref{f2}. In our simulation, we use the following parameters: the detection efficiency is $\eta_{d}=43\%$, the dark count rate of each pulse is $p_{dark}=1\times10^{-7}$, and the intrinsic loss coefficient of standard telecom fibre is $\alpha=0.2$dB/km. These values are adopted from the 200km MDIQKD experiment data\cite{Tang2014MDI200km}.
We also set the misalignment error rate to $e_{d}=0.5\%$, the efficiency of error correction is $f=1.16$. For comparison, we also give the secure key rates of BB84 QKD protocol\cite{Lo:2005:Decoy}, four-state SARG04 QKD protocol\cite{Fung2006SARG04}, and the RRDPS QKD protocol\cite{sasaki2014practical} with the case of infinite decoy states. As shown in Fig. \ref{f2}(a), the secure transmission distance of the six-state SARG04 QKD protocol is more than 270km, farther than the four-state SARG04 QKD protocol because of the higher bit error rate threshold in the six-state SARG04 QKD protocol.
The case of finite decoy states is considered in the Fig. \ref{f2}(b). By exploiting one weak decoy state and vacuum state, one can extract the secure key from single-photon component (see Methods).
However, the secure key rate and secure transmission distance of the six-state and four-state SARG04 QKD protocol are all smaller than those of BB84 protocol since the bit error rate of BB84 protocol is small and the efficiency of basis shift is high\cite{Fung2006SARG04}.
Meanwhile, since the security of RRDPS QKD does not rely on signal disturbance monitoring, in our case where $e_{d}$ is low, the secure key of RRDPS QKD is much lower than qubit-based QKD protocols.

\section*{Discussion}
For each QKD protocol, how to extract as much secure key as possible is a critical task. Here, we present the exact relations between the phase error and bit error as well as the mutual information parameters with single-photon, two-photon, three-photon, and four-photon sources.
Through restricting the mutual information, we have obtained higher bit error rate thresholds of three-photon, four-photon six-state SARG04 and two-photon four-state SARG04 QKD protocol. In the quantum digital signature protocol with $k+1$-participant\cite{gottesman2001quantum,Yin:2016:QDS} (one signer and $k$ recipients), the signer will prepares $k$ copies of quantum states and send a copy of quantum states to each recipient through the insecure quantum channel. To guarantee the security against the forgery attack of untruthful recipient, the honest recipient needs to estimate the information leak of his received quantum states, which will correlate to the phase error rate of QKD with $k$-photon sources.  The security analysis of the four-state and six-state SARG04 QKD protocol with two-photon sources has been used for the three-participants quantum digital signature\cite{Yin:2016:QDS}. Similarly, we can expect that the security analysis of the six-state SARG04 QKD protocol with three and four-photon sources can also be used for the four-participant and five-participant quantum digital signature.

\section*{Methods}

\textbf{The six-state SARG04 protocol with single-photon source.}
We consider the following four orthogonal Bell states
\begin{equation} \label{qubit}
\begin{aligned}
\ket{\Phi^{\pm}}=\frac{1}{\sqrt{2}}(\ket{00}\pm\ket{11}),\\
\ket{\Psi^{\pm}}=\frac{1}{\sqrt{2}}(\ket{01}\pm\ket{10}).\\
\end{aligned}
\end{equation}
Alice prepares the initial quantum state $\ket{\Psi_{AB}^{(1)}}=\frac{1}{\sqrt{2}}(\ket{0_z}_{A}\ket{\varphi_0}_{B}+\ket{1_z}_{A}\ket{\varphi_1}_{B})$. If Eve performs no attacks and Bob does a successful filtering operation, the quantum state shared by Alice and Bob can be given by
\begin{equation} \label{eq1}
\begin{aligned}
\ket{\Psi}&=F\ket{\Psi_{AB}^{(1)}}=\frac{1}{2\sqrt{2}}(\ket{0_z}_{A}\ket{0_{z}}_{B}+\ket{1_z}_{A}\ket{1_{z}}_{B})=\frac{1}{2}\ket{\Phi^{+}}.
\end{aligned}
\end{equation}
Here, we consider that Eve can perform the most general attack on all qubits transmitted through the insecure quantum channel. By tracing out all other qubits, we can focus on one qubit state. Let $\rho_{\textrm{qubit}}$ represent a pair of qubit states that Alice and Bob share after Eve's attack, which can be given by
\begin{equation} \label{qubit}
\begin{aligned}
\rho_{\textrm{qubit}}&=\sum_{l,k}\hat{P}\left[I_{A}\otimes(FR_{B}^{-k}T_{l_{B}}^{-1}E_{B}T_{l_{B}}R_{B}^{k})\ket{\Psi_{AB}^{(1)}}\right],\\
\end{aligned}
\end{equation}
where $l\in\{0,1,2\}$, $k\in\{0,1,2,3\}$, and
\begin{equation} \label{qubit1}
\begin{aligned}
E_{B}=\left( \begin{array}{cc} a_{1} & a_{2}\\ a_{3} & a_{4} \end{array} \right),~~\hat{P}(X\ket{\Psi})=X\ket{\Psi}\bra{\Psi}X^{\dagger}.
\end{aligned}
\end{equation}
Here, $E_{B}$ is a $2\times2$ matrix representing Eve's operations on the single-photon qubit. Meanwhile, any quantum state in the form of a bipartite density matrix can be expressed by the Bell-basis diagonal states. From Eq. \eqref{eq1}, we can see
that the Bell state $\ket{\Phi^{+}}$ is a reference state. Thus, we have
\begin{equation} \label{probability1}
\begin{aligned}
P_{X}=\textrm{Tr}[\rho_{\textrm{qubit}}\ket{\Psi^{+}}\bra{\Psi^{+}}],~~P_{Z}=\textrm{Tr}[\rho_{\textrm{qubit}}\ket{\Phi^{-}}\bra{\Phi^{-}}],~~P_{Y}=\textrm{Tr}[\rho_{\textrm{qubit}}\ket{\Psi^{-}}\bra{\Psi^{-}}],\\
\end{aligned}
\end{equation}
representing the probabilities of only bit flip, only phase shift, both bit flip and phase shift, respectively.
Therefore, the probabilities of bit flip and phase shift can be given by
\begin{equation} \label{probability}
\begin{aligned}
p_{\textrm{bit}}=P_{X}+P_{Y},~~p_{\textrm{ph}}=P_{Z}+P_{Y}.\\
\end{aligned}
\end{equation}
Let $p_{\textrm{fil}}=\textrm{Tr}[\rho_{\textrm{qubit}}]$ represent the trace value of state $\rho_{\textrm{qubit}}$. One can clearly see that  $p_{\textrm{fil}}=\vec{a}^{*}A_{\textrm{fil}}\vec{a}^{T}$, $p_{\textrm{bit}}=\vec{a}^{*}A_{\textrm{bit}}\vec{a}^{T}$, $p_{\textrm{ph}}=\vec{a}^{*}A_{\textrm{ph}}\vec{a}^{T}$, and $P_{Y}=\vec{a}^{*}A_{Y}\vec{a}^{T}$, where $A_{\textrm{fil}}$, $A_{\textrm{bit}}$, $A_{\textrm{ph}}$, and $A_{Y}$ are all $4\times 4$  diagonal matrices, and $\vec{a}=(a_{1}~ a_{2} ~a_{3}~a_{4})$ is a $1\times 4$ vector. If $xA_{\textrm{bit}}+yA_{\textrm{fil}}-A_{\textrm{ph}}\geq 0$ is a positive semi-definite matrix, $xp_{\textrm{bit}}+yp_{\textrm{fil}}\geq p_{\textrm{ph}}$ will always be satisfied.
If $ p_{\textrm{ph}}\leq xp_{\textrm{bit}}+yp_{\textrm{fil}}$ holds, then $e_{p}\leq xe_{b}+y$ becomes exponentially reliable as the number of successfully filtered states increases\cite{Tamaki:2006:SARGtwophoton}.
By using the same argument, if $xA_{\textrm{bit}} \leq A_{Y}\leq yA_{\textrm{bit}}$ holds, then $xe_{b}\leq a\leq ye_{b}$.
The conditional Shannon entropy function can be given by
\begin{equation} \label{mutual information}
\begin{aligned}
H(e_{p}|e_{b})=&-(1+a-e_{b}-e_{p})\log_{2}\frac{1+a-e_{b}-e_{p}}{1-e_{b}}\\
&-(e_{p}-a)\log_{2}\frac{e_{p}-a}{1-e_{b}}-(e_{b}-a)\log_{2}\frac{e_{b}-a}{e_{b}}-a\log_{2}\frac{a}{e_{b}}.
\end{aligned}
\end{equation}

\textbf{The six-state SARG04 protocol with multiphoton sources.} In the case of two-photon, for each quantum state prepared by Alice, the density matrix of quantum state shared by Alice and Bob after Eve's attack can be given by
\begin{equation} \label{qubit}
\begin{aligned}
\rho_{\textrm{qubit}}&=\sum_{l,k,u}\hat{P}[I_{A}\otimes(FR_{B}^{-k}T_{l_{B}}^{-1}E_{B}^{(u)}T_{l_{B}}R_{B}^{k})\ket{\xi^{(l,k,u)}}],\\
\ket{\xi^{(l,k,u)}}&=\frac{1}{\sqrt{2}}[\bra{u_{x}}T_{l}R^{k}\ket{\varphi_{0}}\ket{0_{z}}_{A}\ket{\varphi_{0}}_{B}+\bra{u_{x}}T_{l}R^{k}\ket{\varphi_{1}}\ket{1_{z}}_{A}\ket{\varphi_{1}}_{B}],\\
\end{aligned}
\end{equation}
where $l\in\{0,1,2\}$, $k\in\{0,1,2,3\}$, $u\in\{0,1\}$ and
\begin{equation} \label{qubit}
\begin{aligned}
E_{B}^{(0)}=\bra{0_{x}}E_{B}\ket{0_{x}}=\left( \begin{array}{cc} a_{1} & a_{2}\\ a_{3} & a_{4} \end{array} \right),~~E_{B}^{(1)}=\bra{0_{x}}E_{B}\ket{1_{x}}=\left( \begin{array}{cc} a_{5} & a_{6}\\ a_{7} & a_{8} \end{array} \right),\\
\end{aligned}
\end{equation}
$E_{B}$ is a $4\times4$ matrix which depends on Eve's operation on the two-photon qubit and we can safely assume that the final state of Eve's system is a particular state $\ket{0_{x}}$.
$A_{\textrm{fil}}$, $A_{\textrm{bit}}$, $A_{\textrm{ph}}$, and $A_{Y}$ are $8\times 8$  diagonal matrices,  $\vec{a}=(a_{1}~ a_{2} ~a_{3}~a_{4}~a_{5}~ a_{6} ~a_{7}~a_{8})$ is a $1\times 8$ vector.

In the case of three-photon, for each quantum state prepared by Alice, the density matrix of quantum state shared by Alice and Bob after Eve's attack can be given by
\begin{equation} \label{qubit}
\begin{aligned}
\rho_{\textrm{qubit}}=&\sum_{l,k,u,v}\hat{P}[I_{A}\otimes(FR_{B}^{-k}T_{l_{B}}^{-1}E_{B}^{(u,v)}T_{l_{B}}R_{B}^{k})\ket{\xi^{(l,k,u,v)}}],\\
\ket{\xi^{(l,k,u,v)}}=&\frac{1}{\sqrt{2}}[\bra{u_{x}}T_{l}R^{k}\ket{\varphi_{0}}\bra{v_{x}}T_{l}R^{k}\ket{\varphi_{0}}\ket{0_{z}}_{A}\ket{\varphi_{0}}_{B}\\
&+\bra{u_{x}}T_{l}R^{k}\ket{\varphi_{1}}\bra{v_{x}}T_{l}R^{k}\ket{\varphi_{1}}\ket{1_{z}}_{A}\ket{\varphi_{1}}_{B}],\\
\end{aligned}
\end{equation}
where $l\in\{0,1,2\}$, $k\in\{0,1,2,3\}$, $u\in\{0,1\}$, $v\in\{0,1\}$ and
\begin{equation} \label{qubit}
\begin{aligned}
E_{B}^{(0,0)}&=\bra{0_{x}0_{x}}E_{B}\ket{0_{x}0_{x}}=\left( \begin{array}{cc} a_{1} & a_{2}\\ a_{3} & a_{4} \end{array} \right),~~E_{B}^{(0,1)}=\bra{0_{x}0_{x}}E_{B}\ket{0_{x}1_{x}}=\left( \begin{array}{cc} a_{5} & a_{6}\\ a_{7} & a_{8} \end{array} \right),\\
E_{B}^{(1,0)}&=\bra{0_{x}0_{x}}E_{B}\ket{1_{x}0_{x}}=\left( \begin{array}{cc} a_{9} & a_{10}\\ a_{11} & a_{12} \end{array} \right),~~E_{B}^{(1,1)}=\bra{0_{x}0_{x}}E_{B}\ket{1_{x}1_{x}}=\left( \begin{array}{cc} a_{13} & a_{14}\\ a_{15} & a_{16} \end{array} \right),\\
\end{aligned}
\end{equation}
$E_{B}$ is a $8\times8$ matrix which depends on Eve's operation on the three-photon qubit.
$A_{\textrm{fil}}$, $A_{\textrm{bit}}$, $A_{\textrm{ph}}$, and $A_{Y}$ are $16\times 16$  diagonal matrices,  $\vec{a}=(a_{1}~ a_{2} ~a_{3}~a_{4}~a_{5}~ a_{6} ~a_{7}~a_{8}~a_{9}~ a_{10} ~a_{11}~a_{12}~a_{13}~ a_{14} ~a_{15}~a_{16})$ is a $1\times 16$ vector.

In the case of four-photon, for each quantum state prepared by Alice, the density matrix of quantum state shared by Alice and Bob after Eve's attack can be given by
\begin{equation} \label{qubit}
\begin{aligned}
\rho_{\textrm{qubit}}=&\sum_{l,k,u,v,s}\hat{P}[I_{A}\otimes(FR_{B}^{-k}T_{l_{B}}^{-1}E_{B}^{(u,v,s)}T_{l_{B}}R_{B}^{k})\ket{\xi^{(l,k,u,v,s)}}],\\
\ket{\xi^{(l,k,u,v,s)}}=&\frac{1}{\sqrt{2}}[\bra{u_{x}}T_{l}R^{k}\ket{\varphi_{0}}\bra{v_{x}}T_{l}R^{k}\ket{\varphi_{0}}\bra{s_{x}}T_{l}R^{k}\ket{\varphi_{0}}\ket{0_{z}}_{A}\ket{\varphi_{0}}_{B}\\
&+\bra{u_{x}}T_{l}R^{k}\ket{\varphi_{1}}\bra{v_{x}}T_{l}R^{k}\ket{\varphi_{1}}\bra{s_{x}}T_{l}R^{k}\ket{\varphi_{1}}\ket{1_{z}}_{A}\ket{\varphi_{1}}_{B}],\\
\end{aligned}
\end{equation}
where $l\in\{0,1,2\}$, $k\in\{0,1,2,3\}$, $u\in\{0,1\}$, $v\in\{0,1\}$, $s\in\{0,1\}$ and
\begin{equation} \label{qubit}
\begin{aligned}
E_{B}^{(0,0,0)}&=\bra{0_{x}0_{x}0_{x}}E_{B}\ket{0_{x}0_{x}0_{x}}=\left( \begin{array}{cc} a_{1} & a_{2}\\ a_{3} & a_{4} \end{array} \right),\\
E_{B}^{(0,0,1)}&=\bra{0_{x}0_{x}0_{x}}E_{B}\ket{0_{x}0_{x}1_{x}}=\left( \begin{array}{cc} a_{5} & a_{6}\\ a_{7} & a_{8} \end{array} \right),\\
E_{B}^{(0,1,1)}&=\bra{0_{x}0_{x}0_{x}}E_{B}\ket{0_{x}1_{x}1_{x}}=\left( \begin{array}{cc} a_{9} & a_{10}\\ a_{11} & a_{12} \end{array} \right),\\
E_{B}^{(1,1,1)}&=\bra{0_{x}0_{x}0_{x}}E_{B}\ket{1_{x}1_{x}1_{x}}=\left( \begin{array}{cc} a_{13} & a_{14}\\ a_{15} & a_{16} \end{array} \right).\\
\end{aligned}
\end{equation}
Since the photons of Eve's system are identical, considering their symmetry, we have
\begin{equation} \label{qubit}
\begin{aligned}
E_{B}^{(0,0,1)}=E_{B}^{(0,1,0)}=E_{B}^{(1,0,0)},\\
E_{B}^{(0,1,1)}=E_{B}^{(1,0,1)}=E_{B}^{(1,1,0)}.\\
\end{aligned}
\end{equation}
$E_{B}$ is a $16\times16$ matrix which depends on Eve's operation on the four-photon qubit.
$A_{\textrm{fil}}$, $A_{\textrm{bit}}$, $A_{\textrm{ph}}$, and $A_{Y}$ are $16\times 16$  diagonal matrices,  $\vec{a}=(a_{1}~ a_{2} ~a_{3}~a_{4}~a_{5}~ a_{6} ~a_{7}~a_{8}~a_{9}~ a_{10} ~a_{11}~a_{12}~a_{13}~ a_{14} ~a_{15}~a_{16})$ is a $1\times 16$ vector.

\textbf{Decoy state method with weak coherent state sources.}
By using decoy state method\cite{Hwang:2003:Quantum,Lo:2005:Decoy,Wang:2005:Beating}, one can have
\begin{equation} \label{qubit}
\begin{aligned}
Q_{\mu}=\sum_{n=0}^{\infty}e^{-\mu}\frac{\mu^n}{n!}Y_{n},~~
E_{\mu}Q_{\mu}=\sum_{n=0}^{\infty}e^{-\mu}\frac{\mu^n}{n!}e_{b_{n}}Y_{n}, Q_{n}=e^{-\mu}\frac{\mu^n}{n!}Y_{n}.
\end{aligned}
\end{equation}
where $Y_{n}$ is the yield of $n$-photon. In this simulation, we use the case where Eve
does not interfere with the protocol. For the BB84 protocol, the $Y_{n}$ and $e_{b_{n}}$ can be given by\cite{Lo:2005:Decoy}
\begin{equation} \label{eq1}
\begin{aligned}
Y_{n}=\frac{1}{2}[\eta_{n}+(1-\eta_{n})p_{dark}],~~e_{b_{n}}=\frac{\eta_{n}e_{d}+\frac{1}{2}(1-\eta_{n})p_{dark}}{2Y_{n}},
\end{aligned}
\end{equation}
where $\eta_{n}=1-(1-\eta)^n$, $\eta=\eta_{d}\times10^{-\alpha D/10}$ is the channel transmittance, and $D$ is the distance of optical fibre.
For the RRDPS protocol,
the $Y_{n}$ and $e_{b_{n}}$ can be given by\cite{sasaki2014practical,zhang2015round,Yin:2016:Detector}
\begin{equation} \label{eq2}
\begin{aligned}
Y_{n}=\frac{1}{2L}[\eta_{n}+(1-\eta_{n})p_{dark}L],~~e_{b_{n}}=\frac{\eta_{n}e_{d}+\frac{1}{2}(1-\eta_{n})p_{dark}L}{2LY_{n}},
\end{aligned}
\end{equation}
where $L$ is the number of pulses of each block.
For  the four-state SARG04 protocol,
the $Y_{n}$ and $e_{b_{n}}$ can be given by\cite{Fung2006SARG04}
\begin{equation} \label{eq3}
\begin{aligned}
Y_{n}=\frac{1}{2}[\eta_{n}(e_{d}+\frac{1}{2})+(1-\eta_{n})p_{dark}],~~e_{b_{n}}=\frac{\eta_{n}e_{d}+\frac{1}{2}(1-\eta_{n})p_{dark}}{2Y_{n}}.
\end{aligned}
\end{equation}
For  the six-state SARG04 protocol,
the $Y_{n}$ and $e_{b_{n}}$ can be given by\cite{Yin:2016:QDS}
\begin{equation} \label{eq4}
\begin{aligned}
Y_{n}=\frac{1}{3}[\eta_{n}(e_{d}+\frac{1}{2})+(1-\eta_{n})p_{dark}],~~e_{b_{n}}=\frac{\eta_{n}e_{d}+\frac{1}{2}(1-\eta_{n})p_{dark}}{3Y_{n}}.
\end{aligned}
\end{equation}
For the case with infinite decoy states, one can use the Eqs. \eqref{eq1}-\eqref{eq4} to directly calculate the yield and bit error rate of $n$-photon component\cite{Lo:2005:Decoy,Wang:2005:Beating}.
For the case with finite decoy states, we must estimate the lower bound of yield $Y_{n}^{L}$ and the upper bound of bit error rate $e_{b_{n}}^{U}$. One can exploit three intensities $\mu>\nu>0$ to estimate $Y_{1}^{L}$ and $e_{b_{1}}^{U}$,
\begin{equation} \label{eq}
\begin{aligned}
Y_{1}^{L}\geq\frac{\mu}{\mu  \nu -\nu ^2} \left(e^{\nu } Q_{\nu}-e^{\mu }\frac{ \nu ^2  }{\mu ^2}Q_{\mu}+-\frac{\mu ^2-\nu ^2}{\mu ^2}Y_{0}\right),~~
e_{b_{1}}^{U}\leq \frac{e^\nu E_{\nu}Q_{\nu}-Y_{0}/2}{\nu Y_{1}^{L}}.
\end{aligned}
\end{equation}
One can also exploit four intensities\cite{Yin:2016:QDS} $\mu>\nu>\omega>0$ to estimate $Y_{2}^{L}$ and $e_{b_{2}}^{U}$,
\begin{equation} \label{eq}
\begin{aligned}
Y_{2}^{L}\geq \frac{2}{\mu\nu\omega(\mu-\nu)(\mu-\omega)(\nu-\omega)}\Big\{\mu\omega(\mu^2-\omega^2)e^\nu Q_{\nu}-\mu\nu(\mu^2-\nu^2)e^\omega Q_{\omega}\\
-\nu\omega(\nu^2-\omega^2)e^\mu Q_{\mu}+\big[\mu^3(\nu-\omega)+\nu^3(\omega-\mu)+\omega^3(\mu-\nu)\big]Y_{0}\Big\},
\end{aligned}
\end{equation}
\begin{equation} \label{eq}
\begin{aligned}
e_{b_{2}}^{U}\leq \frac{2}{\nu\omega(\nu-\omega) Y_{2}^{L}}\big[\omega e^\nu E_{\nu}Q_{\nu}-\nu e^\omega E_{\omega}Q_{\omega}+(\nu-\omega)Y_{0}/2\big].
\end{aligned}
\end{equation}
For more photon components, the analytical method will become very complex to calculate the yield and bit error rate.
However, the linear programming\cite{Ma:Statistical:2012} is a good method. To estimate the yield and bit error rate of $n$-photon, one can use $n+2$ kinds of intensities.
Since the probability of multiphoton components is very small in the weak coherent state sources, we simply consider the single-photon component contribution using one signal state, one weak decoy state and vacuum state. The intensity of weak decoy state is 0.1 and the intensity of signal state is optimal for each distance.

\noindent\textbf{Acknowledgments}\\
This work has been supported by the Chinese Academy of Sciences and the National Natural Science Foundation of China under Grant No. 61125502.

\noindent\textbf{Author Contributions}\\
H.-L.Y., Y.F. and Z.-B.C. have the main idea. All results are acquired through the
discussion among all authors. All authors contribute to the writing and reviewing of the manuscript.

\noindent\textbf{Additional Information}\\
Competing financial interests: The authors declare no competing financial interests.

\bibliographystyle{naturemag}


\begin{thebibliography}{10}
\expandafter\ifx\csname url\endcsname\relax
  \def\url#1{\texttt{#1}}\fi
\expandafter\ifx\csname urlprefix\endcsname\relax\def\urlprefix{URL }\fi
\providecommand{\bibinfo}[2]{#2}
\providecommand{\eprint}[2][]{\url{#2}}

\bibitem{BB_84}
\bibinfo{author}{Bennett, C.~H.} \& \bibinfo{author}{Brassard, G.}
\newblock \bibinfo{title}{Quantum cryptography: Public key distribution and
  coin tossing}.
\newblock In \emph{\bibinfo{booktitle}{Proceedings of IEEE International
  Conference on Computers, Systems, and Signal Processing}},
  \bibinfo{pages}{175--179} (\bibinfo{organization}{IEEE, New York},
  \bibinfo{year}{1984}).

\bibitem{e91quantum}
\bibinfo{author}{Ekert, A.~K.}
\newblock \bibinfo{title}{Quantum cryptography based on Bell's theorem}.
\newblock \emph{\bibinfo{journal}{Phys. Rev. Lett.}}
  \textbf{\bibinfo{volume}{67}}, \bibinfo{pages}{661--663}
  (\bibinfo{year}{1991}).

\bibitem{lo1999unconditional}
\bibinfo{author}{Lo, H.-K.} \& \bibinfo{author}{Chau, H.~F.}
\newblock \bibinfo{title}{Unconditional security of quantum key distribution
  over arbitrarily long distances}.
\newblock \emph{\bibinfo{journal}{Science}} \textbf{\bibinfo{volume}{283}},
  \bibinfo{pages}{2050--2056} (\bibinfo{year}{1999}).

\bibitem{shor2000simple}
\bibinfo{author}{Shor, P.~W.} \& \bibinfo{author}{Preskill, J.}
\newblock \bibinfo{title}{Simple proof of security of the BB84 quantum key
  distribution protocol}.
\newblock \emph{\bibinfo{journal}{Phys. Rev. Lett.}}
  \textbf{\bibinfo{volume}{85}}, \bibinfo{pages}{441} (\bibinfo{year}{2000}).

\bibitem{Kraus:2005:Lower}
\bibinfo{author}{Kraus, B.}, \bibinfo{author}{Gisin, N.} \&
  \bibinfo{author}{Renner, R.}
\newblock \bibinfo{title}{Lower and upper bounds on the secret-key rate for
  quantum key distribution protocols using one-way classical communication}.
\newblock \emph{\bibinfo{journal}{Phys. Rev. Lett.}}
  \textbf{\bibinfo{volume}{95}}, \bibinfo{pages}{080501}
  (\bibinfo{year}{2005}).

\bibitem{Tomamichel:2011}
\bibinfo{author}{Tomamichel, M.} \& \bibinfo{author}{Renner, R.}
\newblock \bibinfo{title}{Uncertainty relation for smooth entropies}.
\newblock \emph{\bibinfo{journal}{Phys. Rev. Lett.}}
  \textbf{\bibinfo{volume}{106}}, \bibinfo{pages}{110506}
  (\bibinfo{year}{2011}).

\bibitem{RevModPhys:02:Gisin}
\bibinfo{author}{Gisin, N.}, \bibinfo{author}{Ribordy, G.},
  \bibinfo{author}{Tittel, W.} \& \bibinfo{author}{Zbinden, H.}
\newblock \bibinfo{title}{Quantum cryptography}.
\newblock \emph{\bibinfo{journal}{Rev. Mod. Phys.}}
  \textbf{\bibinfo{volume}{74}}, \bibinfo{pages}{145--195}
  (\bibinfo{year}{2002}).

\bibitem{wang2007quantum}
\bibinfo{author}{Wang, X.-B.}, \bibinfo{author}{Hiroshima, T.},
  \bibinfo{author}{Tomita, A.} \& \bibinfo{author}{Hayashi, M.}
\newblock \bibinfo{title}{Quantum information with gaussian states}.
\newblock \emph{\bibinfo{journal}{Phys. Rep.}} \textbf{\bibinfo{volume}{448}},
  \bibinfo{pages}{1--111} (\bibinfo{year}{2007}).

\bibitem{RevModPhys:09:Scarani}
\bibinfo{author}{Scarani, V.} \emph{et~al.}
\newblock \bibinfo{title}{The security of practical quantum key distribution}.
\newblock \emph{\bibinfo{journal}{Rev. Mod. Phys.}}
  \textbf{\bibinfo{volume}{81}}, \bibinfo{pages}{1301--1350}
  (\bibinfo{year}{2009}).

\bibitem{Weedbrook:2012:CV}
\bibinfo{author}{Weedbrook, C.} \emph{et~al.}
\newblock \bibinfo{title}{Gaussian quantum information}.
\newblock \emph{\bibinfo{journal}{Rev. Mod. Phys.}}
  \textbf{\bibinfo{volume}{84}}, \bibinfo{pages}{621--669}
  (\bibinfo{year}{2012}).

\bibitem{lo2014secureQKD}
\bibinfo{author}{Lo, H.-K.}, \bibinfo{author}{Curty, M.} \&
  \bibinfo{author}{Tamaki, K.}
\newblock \bibinfo{title}{Secure quantum key distribution}.
\newblock \emph{\bibinfo{journal}{Nature Photon.}}
  \textbf{\bibinfo{volume}{8}}, \bibinfo{pages}{595--604}
  (\bibinfo{year}{2014}).

\bibitem{ekert:2014:ultimate}
\bibinfo{author}{Ekert, A.} \& \bibinfo{author}{Renner, R.}
\newblock \bibinfo{title}{The ultimate physical limits of privacy}.
\newblock \emph{\bibinfo{journal}{Nature}} \textbf{\bibinfo{volume}{507}},
  \bibinfo{pages}{443--447} (\bibinfo{year}{2014}).

\bibitem{takesue2007quantum}
\bibinfo{author}{Takesue, H.} \emph{et~al.}
\newblock \bibinfo{title}{Quantum key distribution over a 40-db channel loss
  using superconducting single-photon detectors}.
\newblock \emph{\bibinfo{journal}{Nature Photon.}}
  \textbf{\bibinfo{volume}{1}}, \bibinfo{pages}{343--348}
  (\bibinfo{year}{2007}).

\bibitem{wang:2012}
\bibinfo{author}{Wang, S.} \emph{et~al.}
\newblock \bibinfo{title}{2 GHz clock quantum key distribution over 260 km of
  standard telecom fiber}.
\newblock \emph{\bibinfo{journal}{Opt. Lett.}} \textbf{\bibinfo{volume}{37}},
  \bibinfo{pages}{1008--1010} (\bibinfo{year}{2012}).

\bibitem{Liu:2013:Experimental}
\bibinfo{author}{Liu, Y.} \emph{et~al.}
\newblock \bibinfo{title}{Experimental measurement-device-independent quantum
  key distribution}.
\newblock \emph{\bibinfo{journal}{Phys. Rev. Lett.}}
  \textbf{\bibinfo{volume}{111}}, \bibinfo{pages}{130502}
  (\bibinfo{year}{2013}).

\bibitem{Tang2014MDI200km}
\bibinfo{author}{Tang, Y.-L.} \emph{et~al.}
\newblock \bibinfo{title}{Measurement-device-independent quantum key
  distribution over 200 km}.
\newblock \emph{\bibinfo{journal}{Phys. Rev. Lett.}}
  \textbf{\bibinfo{volume}{113}}, \bibinfo{pages}{190501}
  (\bibinfo{year}{2014}).

\bibitem{korzh2015NatPhot307km}
\bibinfo{author}{Korzh, B.} \emph{et~al.}
\newblock \bibinfo{title}{Provably secure and practical quantum key
  distribution over 307 km of optical fibre}.
\newblock \emph{\bibinfo{journal}{Nature Photon.}}
  \textbf{\bibinfo{volume}{9}}, \bibinfo{pages}{163--168}
  (\bibinfo{year}{2015}).

\bibitem{Tang2016MDInet}
\bibinfo{author}{Tang, Y.-L.} \emph{et~al.}
\newblock \bibinfo{title}{Measurement-device-independent quantum key
  distribution over untrustful metropolitan network}.
\newblock \emph{\bibinfo{journal}{Phys. Rev. X}} \textbf{\bibinfo{volume}{6}},
  \bibinfo{pages}{011024} (\bibinfo{year}{2016}).

\bibitem{frohlich2013QNetwork}
\bibinfo{author}{Fr{\"o}hlich, B.} \emph{et~al.}
\newblock \bibinfo{title}{A quantum access network}.
\newblock \emph{\bibinfo{journal}{Nature}} \textbf{\bibinfo{volume}{501}},
  \bibinfo{pages}{69--72} (\bibinfo{year}{2013}).

\bibitem{lucamarini2015PRXTHA}
\bibinfo{author}{Lucamarini, M.} \emph{et~al.}
\newblock \bibinfo{title}{Practical security bounds against the trojan-horse
  attack in quantum key distribution}.
\newblock \emph{\bibinfo{journal}{Phys. Rev. X}} \textbf{\bibinfo{volume}{5}},
  \bibinfo{pages}{031030} (\bibinfo{year}{2015}).

\bibitem{Lydersen:BrightAttack:2010}
\bibinfo{author}{Lydersen, L.} \emph{et~al.}
\newblock \bibinfo{title}{Hacking commercial quantum cryptography systems by
  tailored bright illumination}.
\newblock \emph{\bibinfo{journal}{Nature Photon.}}
  \textbf{\bibinfo{volume}{4}}, \bibinfo{pages}{686--689}
  (\bibinfo{year}{2010}).

\bibitem{Zhao:2007:Quantum}
\bibinfo{author}{Zhao, Y.}, \bibinfo{author}{Fung, C.-H.~F.},
  \bibinfo{author}{Qi, B.}, \bibinfo{author}{Chen, C.} \& \bibinfo{author}{Lo,
  H.-K.}
\newblock \bibinfo{title}{Quantum hacking: Experimental demonstration of
  time-shift attack against practical quantum-key-distribution systems}.
\newblock \emph{\bibinfo{journal}{Phys. Rev. A}} \textbf{\bibinfo{volume}{78}},
  \bibinfo{pages}{042333} (\bibinfo{year}{2008}).

\bibitem{Lo:2012:MDI}
\bibinfo{author}{Lo, H.-K.}, \bibinfo{author}{Curty, M.} \&
  \bibinfo{author}{Qi, B.}
\newblock \bibinfo{title}{Measurement-device-independent quantum key
  distribution}.
\newblock \emph{\bibinfo{journal}{Phys. Rev. Lett.}}
  \textbf{\bibinfo{volume}{108}}, \bibinfo{pages}{130503}
  (\bibinfo{year}{2012}).

\bibitem{Braunstein:2012:Side}
\bibinfo{author}{Braunstein, S.~L.} \& \bibinfo{author}{Pirandola, S.}
\newblock \bibinfo{title}{Side-channel-free quantum key distribution}.
\newblock \emph{\bibinfo{journal}{Phys. Rev. Lett.}}
  \textbf{\bibinfo{volume}{108}}, \bibinfo{pages}{130502}
  (\bibinfo{year}{2012}).

\bibitem{Ma:Statistical:2012}
\bibinfo{author}{Ma, X.}, \bibinfo{author}{Fung, C.-H.~F.} \&
  \bibinfo{author}{Razavi, M.}
\newblock \bibinfo{title}{Statistical fluctuation analysis for
  measurement-device-independent quantum key distribution}.
\newblock \emph{\bibinfo{journal}{Phys. Rev. A}} \textbf{\bibinfo{volume}{86}},
  \bibinfo{pages}{052305} (\bibinfo{year}{2012}).

\bibitem{yin2014long}
\bibinfo{author}{Yin, H.-L.} \emph{et~al.}
\newblock \bibinfo{title}{Long-distance measurement-device-independent quantum
  key distribution with coherent-state superpositions}.
\newblock \emph{\bibinfo{journal}{Opt. Lett.}} \textbf{\bibinfo{volume}{39}},
  \bibinfo{pages}{5451--5454} (\bibinfo{year}{2014}).

\bibitem{curty:2014:finite}
\bibinfo{author}{Curty, M.} \emph{et~al.}
\newblock \bibinfo{title}{Finite-key analysis for
  measurement-device-independent quantum key distribution}.
\newblock \emph{\bibinfo{journal}{Nature Commun.}}
  \textbf{\bibinfo{volume}{5}}, \bibinfo{pages}{3732} (\bibinfo{year}{2014}).

\bibitem{wang2014simulating}
\bibinfo{author}{Wang, Q.} \& \bibinfo{author}{Wang, X.-B.}
\newblock \bibinfo{title}{Simulating of the measurement-device independent
  quantum key distribution with phase randomized general sources}.
\newblock \emph{\bibinfo{journal}{Sci. Rep.}} \textbf{\bibinfo{volume}{4}},
  \bibinfo{pages}{4612} (\bibinfo{year}{2014}).

\bibitem{yin2014measurement}
\bibinfo{author}{Yin, H.-L.} \emph{et~al.}
\newblock \bibinfo{title}{Measurement-device-independent quantum key
  distribution based on Bell's inequality}.
\newblock \emph{\bibinfo{journal}{arXiv:1407.7375}} .

\bibitem{zhou2015making}
\bibinfo{author}{Zhou, Y.-H.}, \bibinfo{author}{Yu, Z.-W.} \&
  \bibinfo{author}{Wang, X.-B.}
\newblock \bibinfo{title}{Making the decoy-state measurement-device-independent
  quantum key distribution practically useful}.
\newblock \emph{\bibinfo{journal}{Phys. Rev. A}} \textbf{\bibinfo{volume}{93}},
  \bibinfo{pages}{042324} (\bibinfo{year}{2016}).

\bibitem{Xu:2015:Measurement}
\bibinfo{author}{Xu, F.}
\newblock \bibinfo{title}{Measurement-device-independent quantum communication
  with an untrusted source}.
\newblock \emph{\bibinfo{journal}{Phys. Rev. A}} \textbf{\bibinfo{volume}{92}},
  \bibinfo{pages}{012333} (\bibinfo{year}{2015}).

\bibitem{Fu:2016:Long}
\bibinfo{author}{Fu, Y.}, \bibinfo{author}{Yin, H.-L.}, \bibinfo{author}{Chen,
  T.-Y.} \& \bibinfo{author}{Chen, Z.-B.}
\newblock \bibinfo{title}{Long-distance measurement-device-independent
  multiparty quantum communication}.
\newblock \emph{\bibinfo{journal}{Phys. Rev. Lett.}}
  \textbf{\bibinfo{volume}{114}}, \bibinfo{pages}{090501}
  (\bibinfo{year}{2015}).

\bibitem{Acin:2007:DIQKD}
\bibinfo{author}{Ac\'{i}n, A.} \emph{et~al.}
\newblock \bibinfo{title}{Device-independent security of quantum cryptography
  against collective attacks}.
\newblock \emph{\bibinfo{journal}{Phys. Rev. Lett.}}
  \textbf{\bibinfo{volume}{98}}, \bibinfo{pages}{230501}
  (\bibinfo{year}{2007}).

\bibitem{Gisin2010DIQKD}
\bibinfo{author}{Gisin, N.}, \bibinfo{author}{Pironio, S.} \&
  \bibinfo{author}{Sangouard, N.}
\newblock \bibinfo{title}{Proposal for implementing device-independent quantum
  key distribution based on a heralded qubit amplifier}.
\newblock \emph{\bibinfo{journal}{Phys. Rev. Lett.}}
  \textbf{\bibinfo{volume}{105}}, \bibinfo{pages}{070501}
  (\bibinfo{year}{2010}).

\bibitem{Vazirani2014FullDIQKD}
\bibinfo{author}{Vazirani, U.} \& \bibinfo{author}{Vidick, T.}
\newblock \bibinfo{title}{Fully device-independent quantum key distribution}.
\newblock \emph{\bibinfo{journal}{Phys. Rev. Lett.}}
  \textbf{\bibinfo{volume}{113}}, \bibinfo{pages}{140501}
  (\bibinfo{year}{2014}).

\bibitem{Agarwal:1991:Agarwal}
\bibinfo{author}{Agarwal, G.~S.} \& \bibinfo{author}{Tara, K.}
\newblock \bibinfo{title}{Nonclassical properties of states generated by the
  excitations on a coherent state}.
\newblock \emph{\bibinfo{journal}{Phys. Rev. A}} \textbf{\bibinfo{volume}{43}},
  \bibinfo{pages}{492--497} (\bibinfo{year}{1991}).

\bibitem{Brassard2000PNS}
\bibinfo{author}{Brassard, G.}, \bibinfo{author}{L\"utkenhaus, N.},
  \bibinfo{author}{Mor, T.} \& \bibinfo{author}{Sanders, B.~C.}
\newblock \bibinfo{title}{Limitations on practical quantum cryptography}.
\newblock \emph{\bibinfo{journal}{Phys. Rev. Lett.}}
  \textbf{\bibinfo{volume}{85}}, \bibinfo{pages}{1330--1333}
  (\bibinfo{year}{2000}).

\bibitem{azuma2015all}
\bibinfo{author}{Azuma, K.}, \bibinfo{author}{Tamaki, K.} \&
  \bibinfo{author}{Munro, W.~J.}
\newblock \bibinfo{title}{All-photonic intercity quantum key distribution}.
\newblock \emph{\bibinfo{journal}{Nature Commun.}}
  \textbf{\bibinfo{volume}{6}}, \bibinfo{pages}{10171} (\bibinfo{year}{2015}).

\bibitem{abruzzo2014measurement}
\bibinfo{author}{Abruzzo, S.}, \bibinfo{author}{Kampermann, H.} \&
  \bibinfo{author}{Bru{\ss}, D.}
\newblock \bibinfo{title}{Measurement-device-independent quantum key
  distribution with quantum memories}.
\newblock \emph{\bibinfo{journal}{Phy. Rev. A}} \textbf{\bibinfo{volume}{89}},
  \bibinfo{pages}{012301} (\bibinfo{year}{2014}).

\bibitem{mirza2013single}
\bibinfo{author}{Mirza, I.~M.}, \bibinfo{author}{van Enk, S.} \&
  \bibinfo{author}{Kimble, H.}
\newblock \bibinfo{title}{Single-photon time-dependent spectra in coupled
  cavity arrays}.
\newblock \emph{\bibinfo{journal}{J. Opt. Soc. Am. B}}
  \textbf{\bibinfo{volume}{30}}, \bibinfo{pages}{2640--2649}
  (\bibinfo{year}{2013}).

\bibitem{wootters1982NoCloning}
\bibinfo{author}{Wootters, W.~K.} \& \bibinfo{author}{Zurek, W.~H.}
\newblock \bibinfo{title}{A single quantum cannot be cloned}.
\newblock \emph{\bibinfo{journal}{Nature}} \textbf{\bibinfo{volume}{299}},
  \bibinfo{pages}{802--803} (\bibinfo{year}{1982}).

\bibitem{Hwang:2003:Quantum}
\bibinfo{author}{Hwang, W.-Y.}
\newblock \bibinfo{title}{Quantum key distribution with high loss: Toward
  global secure communication}.
\newblock \emph{\bibinfo{journal}{Phys. Rev. Lett.}}
  \textbf{\bibinfo{volume}{91}}, \bibinfo{pages}{057901}
  (\bibinfo{year}{2003}).

\bibitem{Lo:2005:Decoy}
\bibinfo{author}{Lo, H.-K.}, \bibinfo{author}{Ma, X.} \& \bibinfo{author}{Chen,
  K.}
\newblock \bibinfo{title}{Decoy state quantum key distribution}.
\newblock \emph{\bibinfo{journal}{Phys. Rev. Lett.}}
  \textbf{\bibinfo{volume}{94}}, \bibinfo{pages}{230504}
  (\bibinfo{year}{2005}).

\bibitem{Wang:2005:Beating}
\bibinfo{author}{Wang, X.-B.}
\newblock \bibinfo{title}{Beating the photon-number-splitting attack in
  practical quantum cryptography}.
\newblock \emph{\bibinfo{journal}{Phys. Rev. Lett.}}
  \textbf{\bibinfo{volume}{94}}, \bibinfo{pages}{230503}
  (\bibinfo{year}{2005}).

\bibitem{peng2007ExpDecoyQKD}
\bibinfo{author}{Peng, C.-Z.} \emph{et~al.}
\newblock \bibinfo{title}{Experimental long-distance decoy-state quantum key
  distribution based on polarization encoding}.
\newblock \emph{\bibinfo{journal}{Phys. Rev. Lett.}}
  \textbf{\bibinfo{volume}{98}}, \bibinfo{pages}{010505}
  (\bibinfo{year}{2007}).

\bibitem{Rosenberg:2007:Long}
\bibinfo{author}{Rosenberg, D.} \emph{et~al.}
\newblock \bibinfo{title}{Long-distance decoy-state quantum key distribution in
  optical fiber}.
\newblock \emph{\bibinfo{journal}{Phys. Rev. Lett.}}
  \textbf{\bibinfo{volume}{98}}, \bibinfo{pages}{010503}
  (\bibinfo{year}{2007}).

\bibitem{Schmitt:2007:Exp}
\bibinfo{author}{Schmitt-Manderbach, T.} \emph{et~al.}
\newblock \bibinfo{title}{Experimental demonstration of free-space decoy-state
  quantum key distribution over 144 km}.
\newblock \emph{\bibinfo{journal}{Phys. Rev. Lett.}}
  \textbf{\bibinfo{volume}{98}}, \bibinfo{pages}{010504}
  (\bibinfo{year}{2007}).

\bibitem{SARG04:2004:Quantum}
\bibinfo{author}{Scarani, V.}, \bibinfo{author}{Ac\'{\i}n, A.},
  \bibinfo{author}{Ribordy, G.} \& \bibinfo{author}{Gisin, N.}
\newblock \bibinfo{title}{Quantum cryptography protocols robust against photon
  number splitting attacks for weak laser pulse implementations}.
\newblock \emph{\bibinfo{journal}{Phys. Rev. Lett.}}
  \textbf{\bibinfo{volume}{92}}, \bibinfo{pages}{057901}
  (\bibinfo{year}{2004}).

\bibitem{acin2004SARG04PRA}
\bibinfo{author}{Ac\'{\i}n, A.}, \bibinfo{author}{Gisin, N.} \&
  \bibinfo{author}{Scarani, V.}
\newblock \bibinfo{title}{Coherent-pulse implementations of quantum
  cryptography protocols resistant to photon-number-splitting attacks}.
\newblock \emph{\bibinfo{journal}{Phys. Rev. A}} \textbf{\bibinfo{volume}{69}},
  \bibinfo{pages}{012309} (\bibinfo{year}{2004}).

\bibitem{Branciard2005SARG04}
\bibinfo{author}{Branciard, C.}, \bibinfo{author}{Gisin, N.},
  \bibinfo{author}{Kraus, B.} \& \bibinfo{author}{Scarani, V.}
\newblock \bibinfo{title}{Security of two quantum cryptography protocols using
  the same four qubit states}.
\newblock \emph{\bibinfo{journal}{Phys. Rev. A}} \textbf{\bibinfo{volume}{72}},
  \bibinfo{pages}{032301} (\bibinfo{year}{2005}).

\bibitem{Tamaki:2006:SARGtwophoton}
\bibinfo{author}{Tamaki, K.} \& \bibinfo{author}{Lo, H.-K.}
\newblock \bibinfo{title}{Unconditionally secure key distillation from
  multiphotons}.
\newblock \emph{\bibinfo{journal}{Phys. Rev. A}} \textbf{\bibinfo{volume}{73}},
  \bibinfo{pages}{010302} (\bibinfo{year}{2006}).

\bibitem{Fung2006SARG04}
\bibinfo{author}{Fung, C.-H.~F.}, \bibinfo{author}{Tamaki, K.} \&
  \bibinfo{author}{Lo, H.-K.}
\newblock \bibinfo{title}{Performance of two quantum-key-distribution
  protocols}.
\newblock \emph{\bibinfo{journal}{Phys. Rev. A}} \textbf{\bibinfo{volume}{73}},
  \bibinfo{pages}{012337} (\bibinfo{year}{2006}).

\bibitem{koashi2005security}
\bibinfo{author}{Koashi, M.}
\newblock \bibinfo{title}{Security of quantum key distribution with discrete
  rotational symmetry}.
\newblock \emph{\bibinfo{journal}{arXiv quant-ph/0507154}} .

\bibitem{fang2010passive}
\bibinfo{author}{Xu, F.-X.}, \bibinfo{author}{Wang, S.}, \bibinfo{author}{Han,
  Z.-F.} \& \bibinfo{author}{Guo, G.-C.}
\newblock \bibinfo{title}{Passive decoy state SARG04 quantum-key-distribution
  with practical photon-number resolving detectors}.
\newblock \emph{\bibinfo{journal}{Chin. Phys. B}}
  \textbf{\bibinfo{volume}{19}}, \bibinfo{pages}{100312}
  (\bibinfo{year}{2010}).

\bibitem{ali2010fiber}
\bibinfo{author}{Ali, S.} \& \bibinfo{author}{Wahiddin, M.}
\newblock \bibinfo{title}{Fiber and free-space practical decoy state qkd for
  both BB84 and SARG04 protocols}.
\newblock \emph{\bibinfo{journal}{Eur. Phys. J. D}}
  \textbf{\bibinfo{volume}{60}}, \bibinfo{pages}{405--410}
  (\bibinfo{year}{2010}).

\bibitem{ali2012practical}
\bibinfo{author}{Ali, S.}, \bibinfo{author}{Mohammed, S.},
  \bibinfo{author}{Chowdhury, M.} \& \bibinfo{author}{Hasan, A.~A.}
\newblock \bibinfo{title}{Practical SARG04 quantum key distribution}.
\newblock \emph{\bibinfo{journal}{Opt. Quant. Electron.}}
  \textbf{\bibinfo{volume}{44}}, \bibinfo{pages}{471--482}
  (\bibinfo{year}{2012}).

\bibitem{stucki2011long}
\bibinfo{author}{Stucki, D.} \emph{et~al.}
\newblock \bibinfo{title}{Long-term performance of the swissquantum quantum key
  distribution network in a field environment}.
\newblock \emph{\bibinfo{journal}{New J. Phys.}} \textbf{\bibinfo{volume}{13}},
  \bibinfo{pages}{123001} (\bibinfo{year}{2011}).

\bibitem{jeong2011effects}
\bibinfo{author}{Jeong, Y.-C.}, \bibinfo{author}{Kim, Y.-S.} \&
  \bibinfo{author}{Kim, Y.-H.}
\newblock \bibinfo{title}{Effects of depolarizing quantum channels on BB84 and
  SARG04 quantum cryptography protocols}.
\newblock \emph{\bibinfo{journal}{Laser Phys.}} \textbf{\bibinfo{volume}{21}},
  \bibinfo{pages}{1438--1442} (\bibinfo{year}{2011}).

\bibitem{mizutani2014measurement}
\bibinfo{author}{Mizutani, A.}, \bibinfo{author}{Tamaki, K.},
  \bibinfo{author}{Ikuta, R.}, \bibinfo{author}{Yamamoto, T.} \&
  \bibinfo{author}{Imoto, N.}
\newblock \bibinfo{title}{Measurement-device-independent quantum key
  distribution for scarani-acin-ribordy-gisin 04 protocol}.
\newblock \emph{\bibinfo{journal}{Sci. Rep.}} \textbf{\bibinfo{volume}{4}},
  \bibinfo{pages}{5236} (\bibinfo{year}{2014}).

\bibitem{Yin:2016:QDS}
\bibinfo{author}{Yin, H.-L.}, \bibinfo{author}{Fu, Y.} \&
  \bibinfo{author}{Chen, Z.-B.}
\newblock \bibinfo{title}{Practical quantum digital signature}.
\newblock \emph{\bibinfo{journal}{Phys. Rev. A}} \textbf{\bibinfo{volume}{93}},
  \bibinfo{pages}{032316} (\bibinfo{year}{2016}).

\bibitem{sasaki2014practical}
\bibinfo{author}{Sasaki, T.}, \bibinfo{author}{Yamamoto, Y.} \&
  \bibinfo{author}{Koashi, M.}
\newblock \bibinfo{title}{Practical quantum key distribution protocol without
  monitoring signal disturbance}.
\newblock \emph{\bibinfo{journal}{Nature}} \textbf{\bibinfo{volume}{509}},
  \bibinfo{pages}{475--478} (\bibinfo{year}{2014}).

\bibitem{zhang2015round}
\bibinfo{author}{Zhang, Z.}, \bibinfo{author}{Yuan, X.}, \bibinfo{author}{Cao,
  Z.} \& \bibinfo{author}{Ma, X.}
\newblock \bibinfo{title}{Round-robin differential-phase-shift quantum key
  distribution}.
\newblock \emph{\bibinfo{journal}{arXiv:1505.02481}} .

\bibitem{Yin:2016:Detector}
\bibinfo{author}{Yin, H.-L.}, \bibinfo{author}{Fu, Y.}, \bibinfo{author}{Mao,
  Y.} \& \bibinfo{author}{Chen, Z.-B.}
\newblock \bibinfo{title}{Detector-decoy quantum key distribution without
  monitoring signal disturbance}.
\newblock \emph{\bibinfo{journal}{Phys. Rev. A}} \textbf{\bibinfo{volume}{93}},
  \bibinfo{pages}{022330} (\bibinfo{year}{2016}).

\bibitem{lo2001proof}
\bibinfo{author}{Lo, H.-K.}
\newblock \bibinfo{title}{Proof of unconditional security of six-state quantum
  key distribution scheme}.
\newblock \emph{\bibinfo{journal}{Quantum Inf. Comput.}}
  \textbf{\bibinfo{volume}{1}}, \bibinfo{pages}{81--94} (\bibinfo{year}{2001}).

\bibitem{Tang:2013:source}
\bibinfo{author}{Tang, Y.-L.} \emph{et~al.}
\newblock \bibinfo{title}{Source attack of decoy-state quantum key distribution
  using phase information}.
\newblock \emph{\bibinfo{journal}{Phys. Rev. A}} \textbf{\bibinfo{volume}{88}},
  \bibinfo{pages}{022308} (\bibinfo{year}{2013}).

\bibitem{gottesman2001quantum}
\bibinfo{author}{Gottesman, D.} \& \bibinfo{author}{Chuang, I.}
\newblock \bibinfo{title}{Quantum digital signatures}.
\newblock \emph{\bibinfo{journal}{arXiv quant-ph/0105032}} .

\end{thebibliography}

\newpage

\newpage

\end{document}